\documentclass[11pt]{amsart}

\usepackage{amsmath}
\usepackage{amssymb}
\usepackage{graphicx}

\usepackage{subfig}
\usepackage{url}

\newtheorem{theorem}{Theorem}[section]

\theoremstyle{definition}

\numberwithin{equation}{section}

\AtBeginDocument{{\noindent\small
This is a preprint of a paper whose final and definite form will be published 
in the International Journal \emph{Pure and Applied Functional Analysis}, 
ISSN 2189-3756 (print), ISSN 2189-3764 (online). 
Submitted 01-Dec-2015; revised 17-Mar-2016; accepted for publication 18-Mar-2016.}
\vspace{9mm}}

\title[Modeling, dynamics and optimal control of Ebola virus spread]{Modeling, 
dynamics and optimal control\\ of Ebola virus spread}

\author[A. Rachah]{Amira Rachah}
\address[A. Rachah]{Universit\'e Paul Sabatier, Institut de Math\'ematiques,\newline
\indent 31062 Toulouse, Cedex 9, France}
\email{{\tt arachah@math.univ-toulouse.fr}}

\author[D. F. M. Torres]{Delfim F. M. Torres}
\address[D. F. M. Torres]{Center for Research and Development in Mathematics and\newline
\indent Applications (CIDMA), Department of Mathematics, University of Aveiro,\newline
\indent 3810-193 Aveiro, Portugal}
\email[corresponding author]{{\tt delfim@ua.pt}}



\begin{document}

\begin{abstract}
We present a mathematical analysis of the early detection of Ebola virus.
The propagation of the virus is analysed by using a Susceptible, Infected, 
Recovered (SIR) model. In order to provide useful predictions about the 
potential transmission of the virus, we analyse and simulate the SIR model 
with vital dynamics, by adding demographic effects and an induced death rate.
Then, we compute the equilibria of the model. The numerical simulations 
confirm the theoretical analysis. Our study describes the 2015 
detection of Ebola virus in Guinea, the parameters of the model 
being identified from the World Health Organization data. Finally, we consider 
an optimal control problem of the propagation of the Ebola virus, minimizing 
the number of infected individuals while taking into account 
the cost of vaccination.
\end{abstract}

\maketitle


\section{Introduction}
\label{sec1}

Ebola virus is currently affecting several African countries, mainly Guinea,
Sierra Leone, and Liberia. Ebola was first discovered in 1976 in the Democratic
Republic of the Congo near the Ebola River, where the disease takes its name
\cite{barraya,edward,joseph}. Since then, Ebola outbreaks have appeared
sporadically in Africa. The virus, previously known as Ebola haemorrhagic fever,
is the deadliest pathogens for humans. The early signs and symptoms of the virus
include a sudden onset of fever and intense weakness and headache. Over time,
symptoms become increasingly severe and include diarrhoea, raised rash, internal
and external bleed in (from nose, mouth, eyes and anus). As the virus spreads
through the body, it damages the immune system and organs
\cite{legrand,anon2,tara2,okwar,anon1}. Ebola virus is transmitted to an initial
human by contact with an infected animal's body fluid. On the other hand,
human-to-human transmission can take place with direct contact (through broken
skin or mucous membranes in, for example, the eyes, nose, or mouth) with blood
or body fluids of a person who is sick with or has died from Ebola. It is also
transmitted indirectly via exposure to objects or environment contaminated
with infected secretions \cite{alton,borio,dowel,peter,tara1}.

Mathematical models are a powerful tool for investigating human infectious
diseases, such as Ebola, contributing to the understanding of the dynamics
of the disease, providing useful predictions about the potential transmission
of the disease and the effectiveness of possible control measures, which can
provide valuable information for public health policy makers
\cite{diekman,gaff,MR2719552,delf}.
Epidemic models date back to the early twentieth century, to the 1927 work
by Kermack and McKendrick, whose  model was used for modelling the plague
and cholera  epidemics. In fact, such epidemic models have provided the
foundation for the best vaccination practices for influenza \cite{longini}
and small pox \cite{kretz}. Currently, the simplest and most commonly 
implemented model in epidemiology is the SIR model. The SIR model consists 
of three compartments: Susceptible individuals $S$,
Infectious individuals $I$, and Recovered individuals $R$ \cite{herbert}.
When analysing a new outbreak, the researchers usually start with the SIR model
to fit the available outbreak data and obtaining estimates for the parameters
of the model \cite{brauer}. This has been the case for the modelling of the
spreading mechanism of the Ebola virus currently affecting several African
countries \cite{MyID:321,MyID:336,MyID:331}. For more complex mathematical models, 
with more than three state variables, see \cite{MR3394468,Area2,MyID:340}.

In our previous works \cite{MyID:321,MyID:331}, we used parameters identified 
from the recent data of the World Health Organization (WHO) to describe the 
behaviour of the virus. Here we focus on the mathematical analysis of the early 
detection of the Ebola virus. In Section~\ref{sec2}, we briefly recall the analysis 
study of the SIR model that we presented in our previous study of the description 
of the behaviour of Ebola virus \cite{MyID:321,MyID:336}. In Section~\ref{sec3},
we add to the basic model of Section~\ref{sec2} the demographic effects, 
in order to provide a description of the virus propagation closer to the reality.
This gives answer to an open question posed in Remark~1 of \cite{MyID:340}
and at the end of \cite{Area2}.
Our aim in studying the model with vital dynamics is to provide useful
predictions about the potential transmission of the virus. We also 
consider an induced death rate for the infected individuals. 
After numerical simulations, in Section~\ref{sec4} we control 
the propagation of the virus in order to minimize the number of infected 
individuals and the cost of vaccination. We end with 
Section~\ref{sec:conc} of conclusions.


\section{Mathematical formulation for the basic SIR model}
\label{sec2}

In this section, we present and briefly discuss the properties
of the system of equations corresponding to the basic SIR
(Susceptible--Infectious--Recovery) model, which has recently
been  used in \cite{MyID:321} to describe the
early detection of Ebola virus in West Africa.
In the formulation of the basic SIR model,
we assume that the population size is constant and
any person who has completely recovered from the virus
acquired permanent immunity. Moreover, we assume that
the virus has a negligibly short incubation period, so that an individual
who contracts the virus becomes infective immediately afterwards.
These assumptions enables us to divide the host population into
three categories,
\begin{itemize}
\item $S(t)$ for susceptible: denotes individuals who are susceptible
to catch the virus, and so might become infectious if exposed;

\item $I(t)$ for Infectious: denotes infectious individuals  who are able
to spread the virus through contact with the susceptible category;

\item $R(t)$ for Recovered: denotes individuals who have immunity to the infection,
and consequently do not affect the transmission dynamics
in any way when in contact with other individuals.
\end{itemize}
The model is described mathematically by the following system
of non-linear differential equations:
\begin{equation}
\label{eq1:SIR}
\begin{cases}
\dfrac{dS(t)}{dt} = -\beta S(t)I(t),\\[0.2cm]
\dfrac{dI(t)}{dt} =  \beta S(t)I(t)- \mu I(t),\\[0.2cm]
\dfrac{dR(t)}{dt} =  \mu I(t),
\end{cases}
\end{equation}
where $\beta >0$ is the infection rate and $\mu >0$ is the recovered rate.
The initial conditions are given by
\begin{equation}
\label{eq4:SIR}
S(0)=S_0>0, \quad I(0)=I_0>0, \quad R(0)=0.
\end{equation}
We can see that $\dfrac{d}{dt} \left[ S(t) + I(t) + R(t) \right] = 0$,
that is, the population size $N$ is constant during the period under study:
\begin{equation}
\label{eq5:SIR}
S(t) + I(t) + R(t)=N
\end{equation}
for any $t\geq0$, which is far from the reality. 


\section{SIR model with vital dynamics and an induced death rate}
\label{sec3}

In the basic SIR model \eqref{eq1:SIR}, we ignore the demographic effects
on the population. In this section, we study a variant of the basic model 
by considering vital dynamics, that is, by adding the birth and death rates 
into the model. Moreover, we increase the death rate of the infectious class 
by considering an induced death rate associated to the infected individuals. 
Such model is new in the Ebola context \cite{MyID:321,MyID:331,MR3356525}.


\subsection{Model formulation}
\label{sec3_subsec1}

\begin{figure}[ht!]
\centering
\includegraphics[scale=0.5]{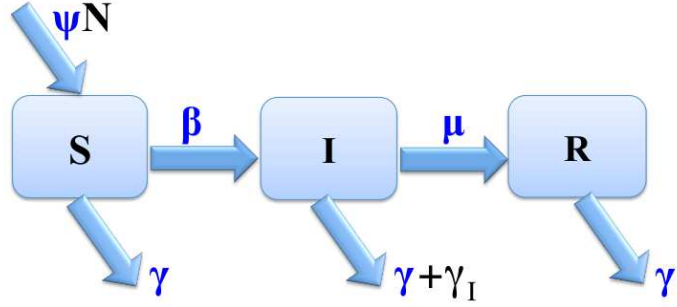}
\caption{Compartment diagram of the SIR model \eqref{SIR_BD2}
with vital dynamics $\psi$ and $\gamma$ 
and an induced death rate $\gamma_{I}$.}
\label{fig:SIR_BD2}
\end{figure}

If we expand the SIR model \eqref{eq1:SIR} by including the demographic
effects, assuming a constant rate of births $\psi$, an equal rate 
of deaths $\gamma$ per unit of time, and an induced death rate 
$\gamma_{I}$, then the mathematical model is described 
by the following system of differential equations:
\begin{equation}
\label{SIR_BD2}
\begin{cases}
\dfrac{dS(t)}{dt} = \psi N - \beta S(t)I(t) - \gamma S(t),\\[0.2cm]
\dfrac{dI(t)}{dt} =  \beta S(t)I(t) - \mu I(t)
- \left(\gamma + \gamma_{I} \right) I(t),\\[0.2cm]
\dfrac{dR(t)}{dt} =  \mu I(t) - \gamma R(t).
\end{cases}
\end{equation}
Figure~\ref{fig:SIR_BD2} shows the compartment diagram 
of the SIR  model \eqref{SIR_BD2} with vital dynamics, that is, 
with demographic (birth and death) effects, and an induced death rate.


\subsection{Analysis of the equilibria}
\label{sec4_subsec1}

Firstly, we start by analysing the equations \eqref{SIR_BD2} of the model 
that serve as the basis for the propagation dynamics of the Ebola virus
with death and birth rates. As we shall see (cf. Theorem~\ref{thm2}),
the dynamics are determined by the basic reproduction number
\begin{equation}
\label{eq:brn_induc}
R_0 := \dfrac{\beta N}{\mu  + \gamma + \gamma_{I}}.
\end{equation}
An equilibrium point $E = (S, I, R)\in \mathrm{R}^{3}_{+}$ 
of \eqref{SIR_BD2} satisfies, by definition, 
\begin{equation}
\label{eq1:eqlbr_induc}
\psi N - \beta SI - \gamma S  = 0,
\end{equation}
\begin{equation}
\label{eq2:eqlbr_induc}
\beta SI - \mu I - \left(\gamma + \gamma_{I}\right) I   = 0,
\end{equation}
\begin{equation}
\label{eq3:eqlbr_induc}
 \mu I - \gamma R = 0.
\end{equation}
There are two biologically meaningful equilibrium points:
if $I=0$, then there is no disease for the population, and
the equilibrium point is called a disease-free equilibrium;
otherwise, if $I>0$, then the equilibrium point is called endemic. 
By adding equations \eqref{eq1:eqlbr_induc} and \eqref{eq2:eqlbr_induc},
we obtain that
\begin{equation*}
\psi N -  \gamma S - (\mu+ \gamma + \gamma_{I})I = 0.
\end{equation*}
Then,
\begin{equation}
\label{eq4:eqlbr_induc}
S  = \dfrac{\psi N  - (\mu+ \gamma + \gamma_{I})I}{\gamma}.
\end{equation}
From \eqref{eq2:eqlbr_induc} and \eqref{eq4:eqlbr_induc}, we get
\begin{equation*}
I\left(\dfrac{\beta\psi N - \gamma \left(\mu + \gamma
+ \gamma_{I}\right)}{\gamma} - \dfrac{\beta\left(\mu
+ \gamma + \gamma_{I}\right)}{\gamma}I\right)  = 0.
\end{equation*}
Therefore, or $I=0$ or
\begin{equation}
\label{eq5:eqlbr_induc}
I  =\dfrac{\beta\psi N - \gamma \left(\mu + \gamma
+ \gamma_{I}\right)}{\beta\left(\mu + \gamma  + \gamma_{I}\right)}.
\end{equation} 
If $I=0$, then from \eqref{eq4:eqlbr_induc} we obtain
that $S=\dfrac{\psi N}{\gamma}$. It follows from \eqref{eq3:eqlbr_induc}
that $R=0$. We just proved that there is a virus free equilibrium $E_1$ given by
\begin{equation}
\label{eq6:SIR_BD_induc}
\lim_{t \to \infty} \left(S(t), I(t), R(t)\right)
= \left(\dfrac{\psi N}{\gamma}, 0, 0\right) =: E_1.
\end{equation}
If \eqref{eq5:eqlbr_induc} holds, then there is another equilibrium with
\begin{equation}
\label{eq7:eqlbr_induc}
I^*  =\dfrac{\beta\psi N - \gamma \left(\mu + \gamma
+ \gamma_{I}\right)}{\beta\left(\mu + \gamma + \gamma_{I}\right)}.
\end{equation}
By substituting \eqref{eq7:eqlbr_induc} into \eqref{eq4:eqlbr_induc},
we obtain that
\begin{equation}
\label{eq8:eqlbr_induc}
S^*  = \dfrac{\mu + \gamma + \gamma_{I}}{\beta} = \dfrac{N}{R_0}
\end{equation}
and, using \eqref{eq3:eqlbr_induc} in \eqref{eq4:eqlbr_induc}, we get
\begin{equation}
\label{eq9:eqlbr_induc}
R^*  = \dfrac{\mu}{\gamma}I^*=\dfrac{\mu}{\gamma}
\left[\dfrac{\beta\psi N - \gamma \left(\mu + \gamma
+ \gamma_{I}\right)}{\beta\left(\mu + \gamma  + \gamma_{I}\right)}\right].
\end{equation}
We just obtained the endemic equilibrium $E_2$ given by
\begin{equation}
\label{eq9:SIR_BD_induc}
\lim_{t \to \infty} \left(S(t), I(t), R(t)\right)
= \left(S^*,I^*,R^*\right) =: E_2,
\end{equation}
where the expressions of $S^*$, $I^*$ and $R^*$ are given by
\eqref{eq7:eqlbr_induc}--\eqref{eq9:eqlbr_induc}.
Next result summarizes what we have obtained so far.

\begin{theorem}
\label{thm2}
Let $R_0$ be the basic reproduction number defined by \eqref{eq:brn_induc}.
If $R_0\leq 1$, then the disease free equilibrium
$E_1=\left(\dfrac{\psi N}{\gamma},0,0\right)$ of the virus is obtained,
which corresponds to the case when the virus dies out (no epidemic).
If $R_0>1$, then the equilibrium $E_2=\left(S^*,I^*,R^*\right)$
of the virus is obtained, in agreement with expressions
\eqref{eq7:eqlbr_induc}--\eqref{eq9:eqlbr_induc},
and the virus is able to invade the population (endemic equilibrium).
\end{theorem}


\subsection{Simulation of the SIR model with demographic 
effects and an induced death rate}
\label{sec4_subsec2}

\begin{figure}[ht!]
\centering
\includegraphics[scale=0.50]{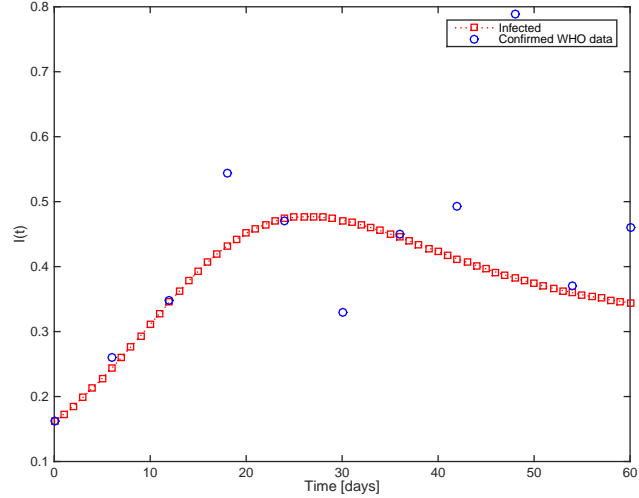}
\caption{Graph of infected obtained from \eqref{SIR_BD2}--\eqref{eq:ic_induced} 
\emph{versus} the real data of confirmed cases for the 2015 Ebola outbreak 
occurred in Guinea between January and March 2015.
\label{I_sml_data_model_induced}}
\end{figure}

\begin{figure}[ht!]
\centering
\subfloat[$S(t)$ with $S(0) = 0.8387$]{%
\label{BD_fig1:SIR_fit}\includegraphics[width=0.5\textwidth]{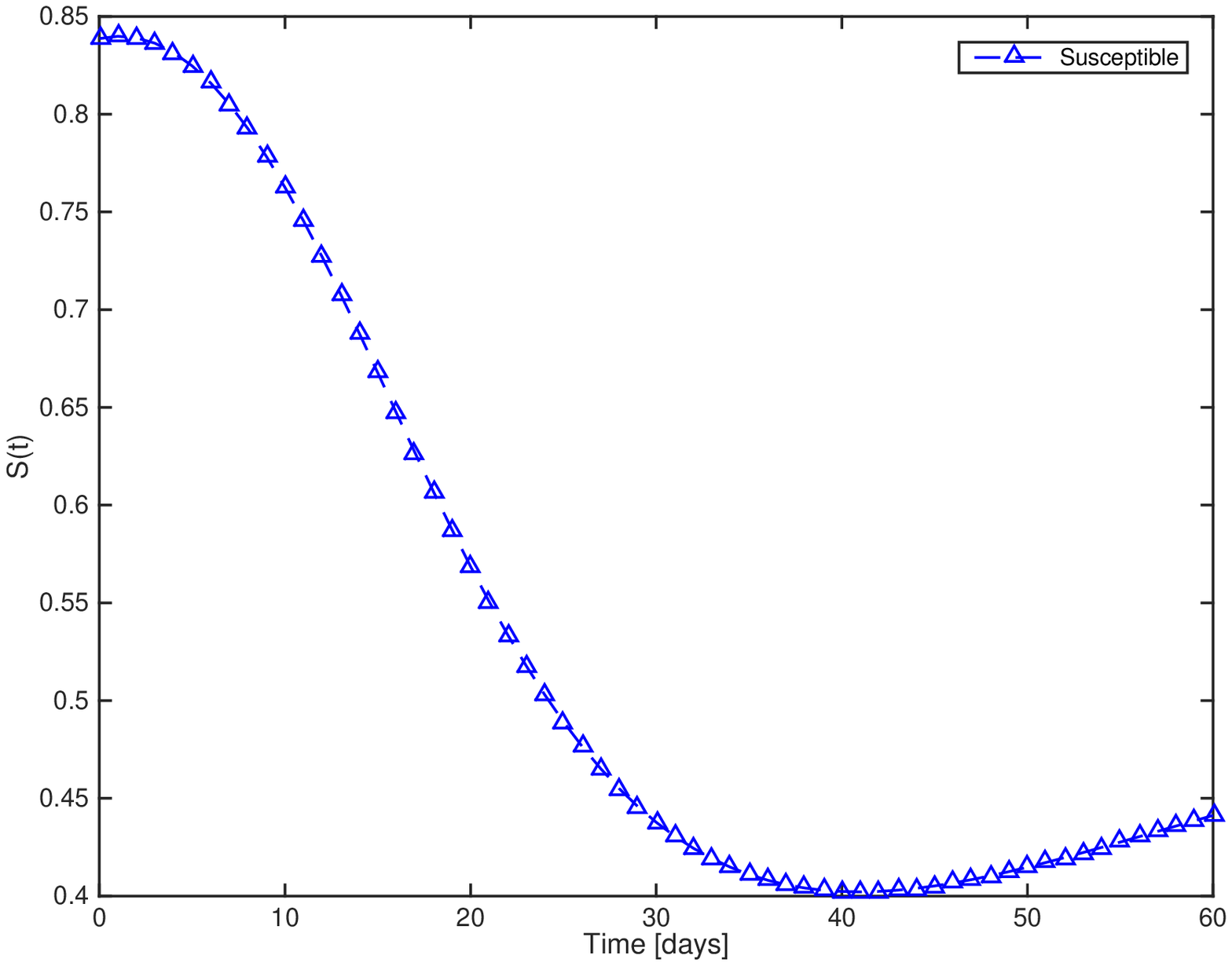}}
\subfloat[$I(t)$ with $I(0) =  0.1613$]{%
\label{BD_fig2:SIR_fit}\includegraphics[width=0.5\textwidth]{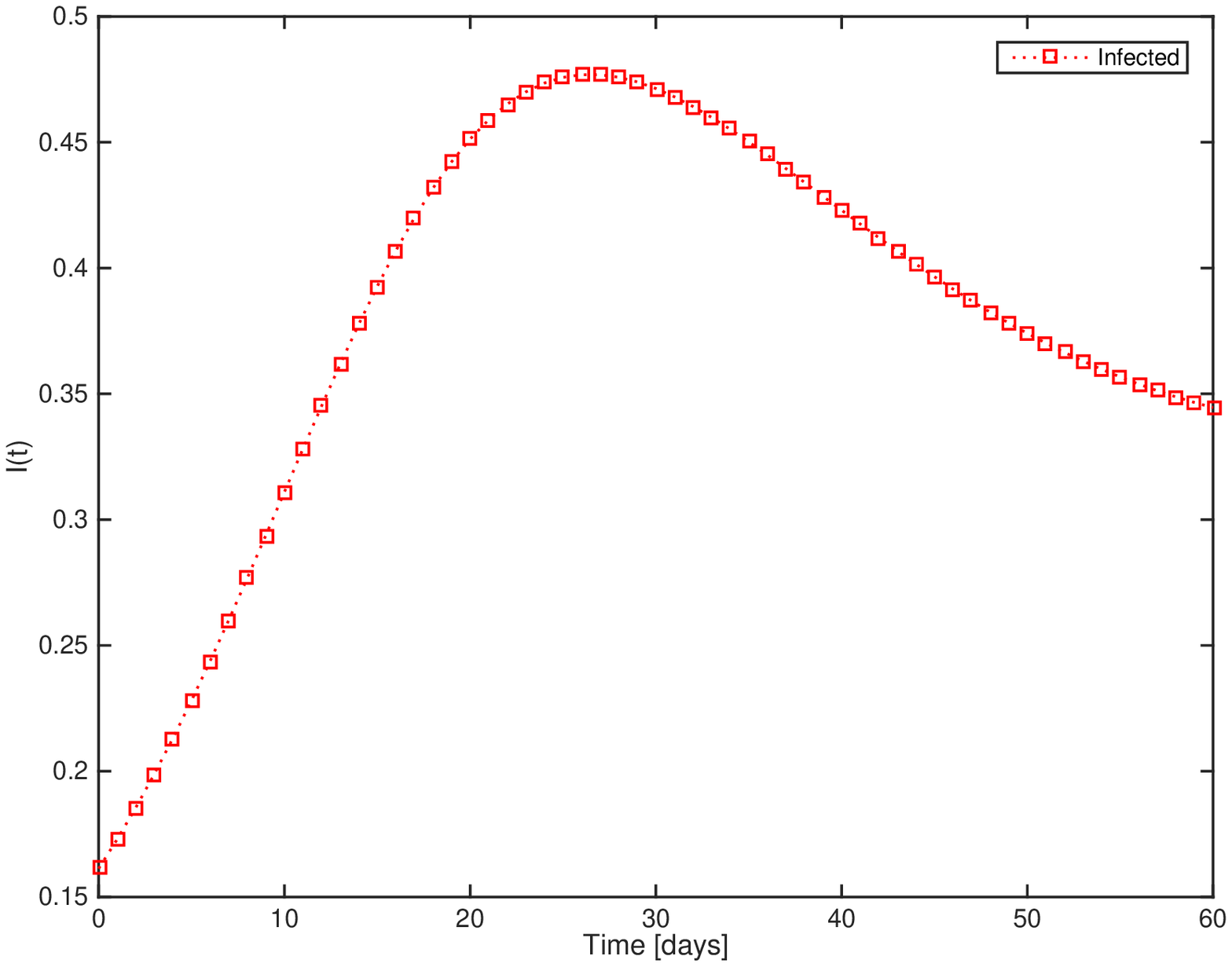}}\\
\subfloat[$R(t)$ with $R(0) = 0$]{%
\label{BD_fig3:SIR_fit}\includegraphics[width=0.5\textwidth]{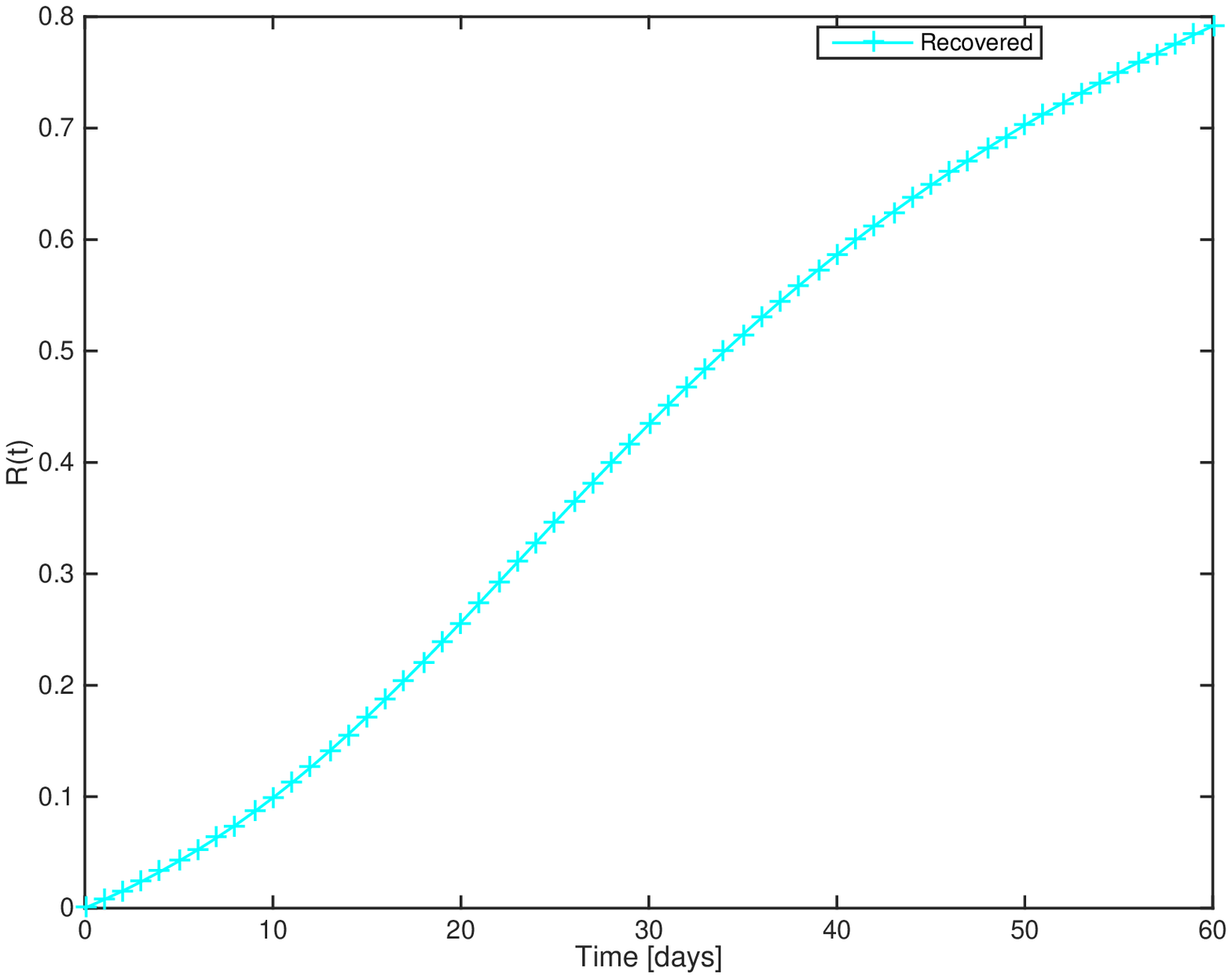}}
\caption{Evolution of individuals in compartments $S(t)$, $I(t)$ and $R(t)$ 
of the SIR model \eqref{SIR_BD2} with vital dynamics and an induced death rate.}
\label{fig:BD:SIR2}
\end{figure}

\begin{figure}[ht!]
\centering
\subfloat[$S(t)$]{
\label{BD_fig1:SIR2}\includegraphics[width=0.49\textwidth]{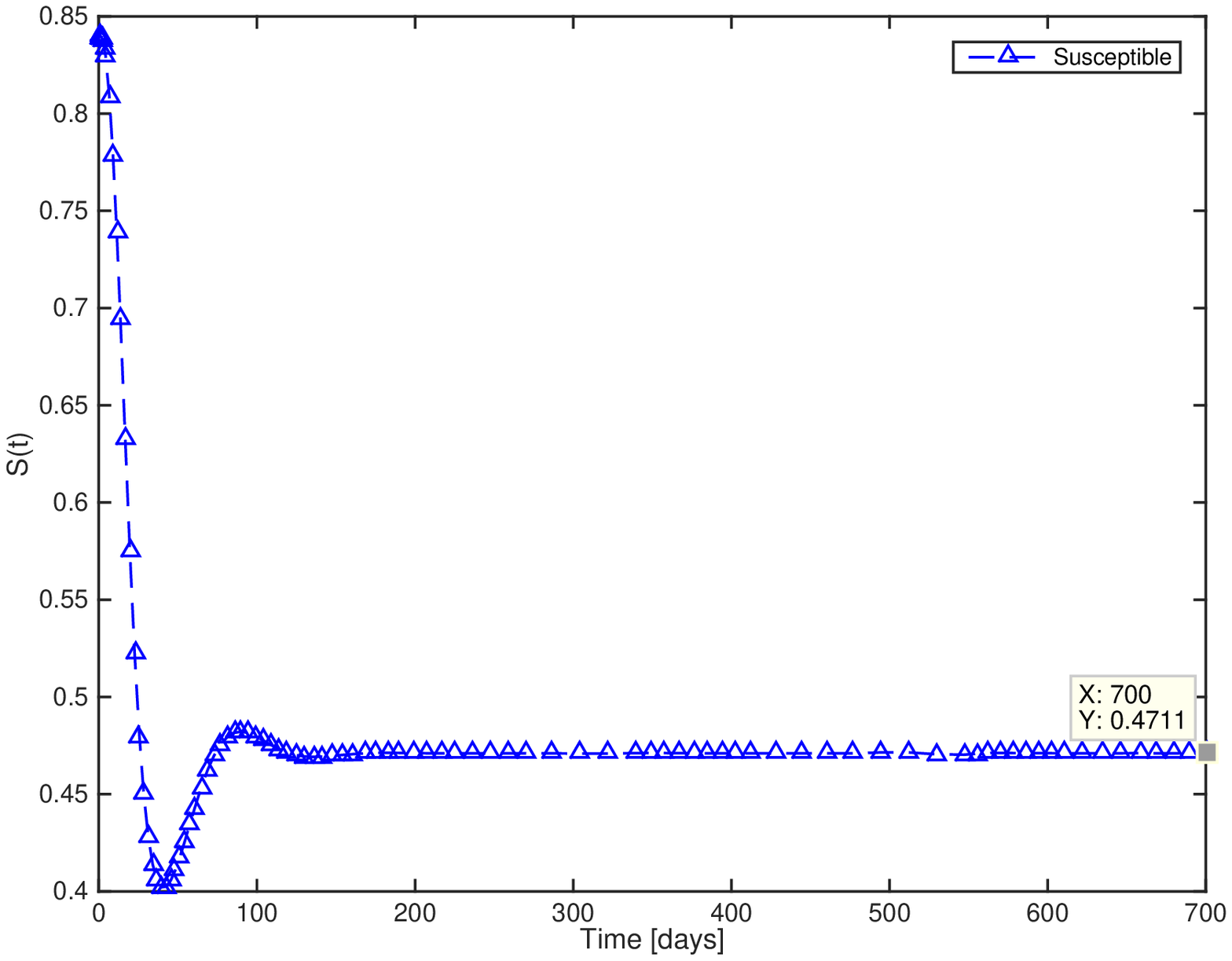}}
\subfloat[$I(t)$]{
\label{BD_fig2:SIR2}\includegraphics[width=0.49\textwidth]{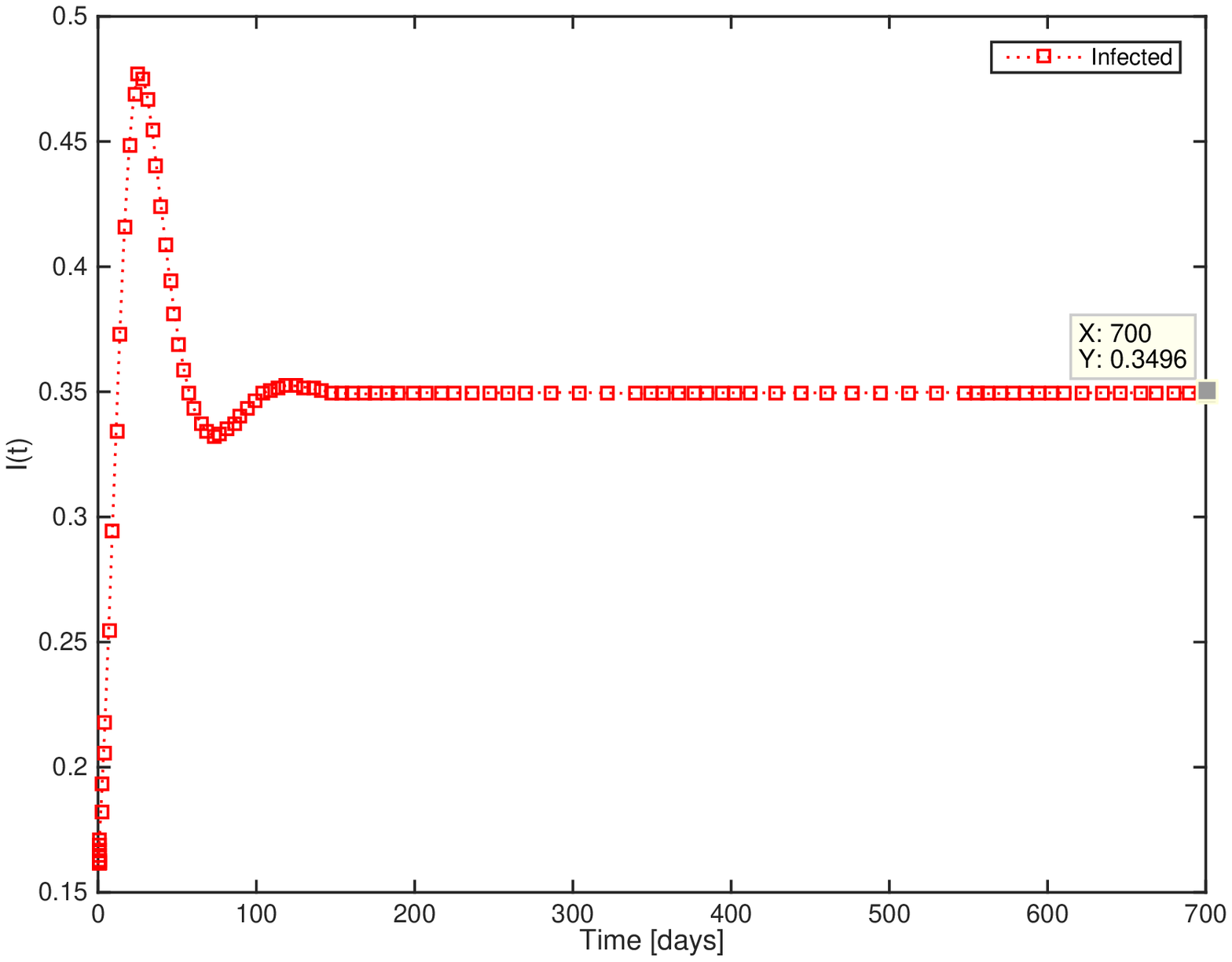}}\\
\subfloat[$R(t)$]{
\label{BD_fig3:SIR2}\includegraphics[width=0.49\textwidth]{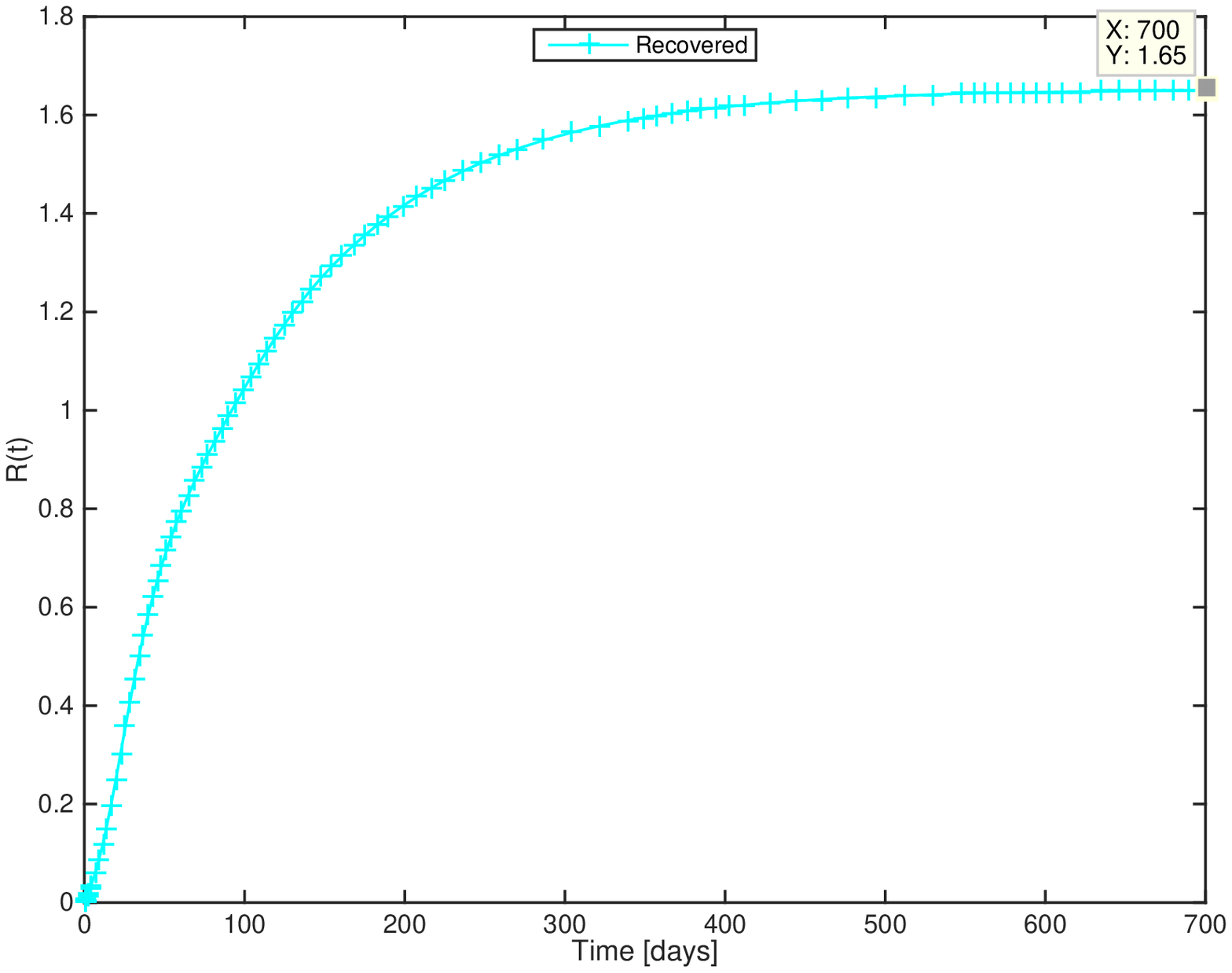}}
\caption{Evolution of individuals in compartments
$S(t)$, $I(t)$ and $R(t)$ of the SIR model \eqref{SIR_BD2}
with vital dynamics and an induced death rate, where the
endemic equilibrium is given by $\left(S^*(t), I^*(t), R^*(t)\right)
=\left(0.47, 0.34, 1.65\right)$.}
\label{fig:BD:SIR2:b}
\end{figure}

We now present a simulation of the model, taking into account
the real outbreak of Ebola virus occurred in Guinea in 2015 
and by using the World Health Organization (WHO) data. 
Precisely, the epidemic data used in our study is borrowed from 
the WHO web site \cite{guinea_who}. The birth rate $\psi=0.03574$ and death rate 
$\gamma=0.00946$ of the model are obtained from the specific statistical 
study of the demographic of Guinea in 2015 \cite{stat_mondial}. The parameters 
$\beta$, $\mu$, $\gamma$ and $\gamma_{I}$ are obtained by identification 
by using the real data of WHO. To estimate the parameters of the model, 
we adapted the initialisation of $I$ with the reported data of WHO by
fitting the real data of confirmed cases of infectious in Guinea.
The result of fitting is shown in Figure~\ref{I_sml_data_model_induced}. 
The comparison between the curve of infectious obtained by our simulation 
and the reported data of confirmed cases by WHO shows that the mathematical 
model \eqref{SIR_BD2} fits well the real data by using $\beta=0.19$ as the  
transmission rate, $\gamma=0.034$ as the infectious rate, 
$\mu=0.0447$ as the recovered rate, and $\gamma_{I}=0.0353$ as the induced 
death rate. By comparing the value of $\gamma_{I}$ with the 
death rate $\gamma$, we remark that $\gamma_{I} = 3.735 \gamma$.
The initial susceptible, infected and recovered populations, 
are given by
\begin{equation}
\label{eq:ic_induced}
S(0)=0.8387, \quad I(0)=0.1613, \quad R(0)=0,
\end{equation}
respectively. The choice of $I(0)$ is in agreement with the data shown in 
Figure~\ref{I_sml_data_model_induced} of the WHO data. By using 
the value of Guinea's population, which is estimated at $P=11780162$ 
in $2015$, and the number of confirmed infectious cases (obtained from WHO), 
the initialization of $I(0)$ corresponds to the number of infected 
divided by the total number of population. Then, in reality, the initial 
number of infected, in the period between January 2015 and March 2015, 
is given by $0.1613 \times P=1900000$, that is, the number of confirmed 
infectious cases represents $16.13\%$ of the total of population.
Figure~\ref{I_sml_data_model_induced} presents the curve of infectious 
individuals $I(t)$ simulated by using \eqref{SIR_BD2} and obtained 
from the WHO real data.
The evolution of susceptible, infected and recovered groups over
time, is shown in Figure~\ref{fig:BD:SIR2}.
We also studied numerically the equilibria, by solving numerically the SIR model 
\eqref{SIR_BD2} with the same parameters and the same initialization. 
Figure~\ref{BD_fig1:SIR2} shows the evolution of the susceptible individuals 
$S(t)$ over time. We see that the oscillations in the numbers of the three 
compartments damp out over time, eventually reaching an equilibrium. 
In our mathematical analysis of the model, we found that the equilibrium $S^*$ 
is computed theoretically by \eqref{eq8:eqlbr_induc}. When we calculate the value 
of this theoretical result, we find $S^*=0.47$, which is equal to the $S^*$ 
computed by the numerical resolution of the model. Figure~\ref{BD_fig2:SIR2} 
shows the evolution of the infected individuals $I(t)$ over time.
We know that the equilibrium $I^*$ is given by \eqref{eq7:eqlbr_induc}. 
When we calculate the value of this theoretical result, we find $I^*=0.34$, 
which agrees with the value computed by the numerical resolution of the model.
Figure~\ref{BD_fig3:SIR2} shows the evolution of the recovered individuals
$R(t)$ over time. The equilibrium $R^*$ is in this case computed theoretically 
by \eqref{eq9:eqlbr_induc}, which coincides the value $R^*=1.65$ computed 
numerically.
The fact that the reached equilibrium $(S^*, I^*, R^*)$ computed theoretically
coincides with the value found by the numerical simulation of the model, 
is a validation of our study of the SIR model with vital dynamics and induced 
death rate, which describes well the currently detection of Ebola virus in Guinea.


\section{Optimal control of the virus under vital dynamics}
\label{sec4}

Nowadays there are several trial vaccinations against Ebola. 
One was already applied in Guinea and seems highly effective
\cite{guinea_who2}. In this section, we present 
a strategy for the control of the virus,
by introducing into the model \eqref{SIR_BD2} a control $u(t)$
representing the vaccination rate at time $t$. The control $u(t)$
is the fraction of susceptible individuals being vaccinated per unit of time,
taking values on the interval $[0, 0.9]$. Then, the mathematical model 
with control is given by the following system of non-linear differential equations:
\begin{equation}
\label{SIR_BD_control}
\begin{cases}
\dfrac{dS(t)}{dt} = \psi N - \beta S(t)I(t) - \gamma S(t)  - u(t) S(t),\\[0.2cm]
\dfrac{dI(t)}{dt} =  \beta S(t)I(t) - \mu I(t) - (\gamma + \gamma_I) I(t),\\[0.2cm]
\dfrac{dR(t)}{dt} =  \mu I(t) - \gamma R(t) +  u(t) S(t).
\end{cases}
\end{equation}
The goal of the strategy is to reduce the infected individuals and the cost
of vaccination. Precisely, the optimal control problem consists of minimizing
the objective functional
\begin{equation}
\label{cost_func}
J(u) = \int_{0}^{t_{end}} \left[I(t) + \dfrac{\tau}{2}u^2(t)\right] dt,
\end{equation}
where $u$ is the control variable, $u(t) \in [0, 0.9]$,
which represents the vaccination rate
at time $t$, and the parameters $\tau$ and $t_{end}$ denote, respectively,
the weight on cost and the duration of the vaccination program.
In the quadratic term of \eqref{cost_func}, $\tau$ is a positive
weight parameter associated with the control $u(t)$
and the square of the control variable reflects the severity
of the side effects of the vaccination. One has
$u \in \mathcal{U}_{ad}$, where
$$
\mathcal{U}_{ad}=\left\{u : u \,  \text{is measurable}, 0
\leq u(t) \leq u_{max}<\infty, \, t\in [0,t_{end}] \right\}
$$
is the admissible control set with $u_{max}=0.9$. The existence 
of an optimal solution follows from Theorem~2.1 of \cite{MR2277889}
(see also the study of the spread of influenza A (H1N1) 
\cite{elhia_exist_control} by using the SIR model).

In our study of the control of the virus, we use the parameters
defined in Section~\ref{sec4_subsec2}. For the numerical simulations
of the optimal control problem, we have used the \textsf{ACADO} solver
\cite{acado}, which is based on a multiple shooting method, including automatic
differentiation and based ultimately on the semidirect multiple shooting algorithm
of Bock and Pitt \cite{acado2}. The \textsf{ACADO} solver comes as a self-contained
public domain software environment, written in \textsf{C++},
for automatic control and dynamic optimization.
Because \eqref{SIR_BD_control} is a nonlinear control system, 
functional \eqref{cost_func} is quadratic in the control 
and linear in the phase variable $I$, it is not clear that the 
numerically found solution through \textsf{ACADO} to our optimal control 
problem gives the global minimum to the functional \eqref{cost_func}, 
or only a local one. Indeed, with \textsf{ACADO} the optimal control problem 
is approximated by a finite dimensional optimization problem, which is then solved 
by techniques from mathematical programming. This only gives a candidate 
for local minimizer. Because of this, we have also used a dynamic programming 
approach and checked our results by solving it with \textsf{BocopHJB}, 
which is a \textsf{C}/\textsf{C++} toolbox for optimal control 
developed since 2014 in the framework of the Inria--Saclay 
initiative for an open source optimal control toolbox, being 
supported by the team ``Commands'' (see \url{http://bocop.org}). 
While \textsf{ACADO} implements a local optimization method,  
the package \textsf{BocopHJB} implements a global optimization method. 
Similarly to the dynamic programming approach, the optimal control 
problem is solved by \textsf{BocopHJB} in two steps: 
first the Hamilton--Jacobi--Bellman equation satisfied 
by the value function of the problem is solved; 
then the optimal trajectory is simulated from 
any chosen initial condition \cite{BocopHJB}.
The obtained results through \textsf{BocopHJB} 
are coincident with those obtained through \textsf{ACADO} and, 
because of that, we claim to have found the global minimum to the problem. 
\begin{figure}[ht!]
\centering
\subfloat[Susceptible $S(t)$]{
\label{cntrlBD_S}\includegraphics[width=0.50\textwidth]{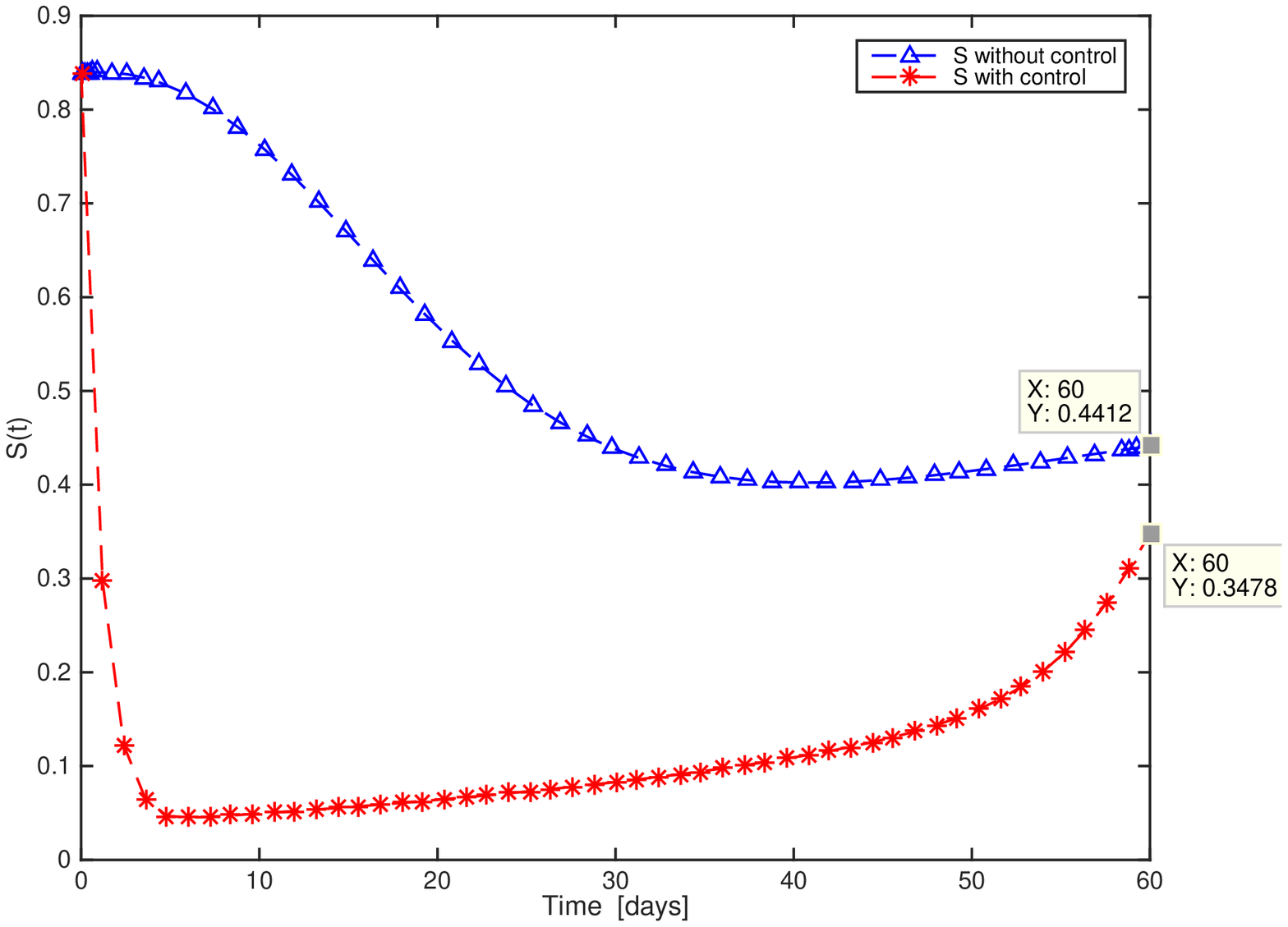}}
\subfloat[Recovered $R(t)$]{
\label{cntrlBD_R}\includegraphics[width=0.50\textwidth]{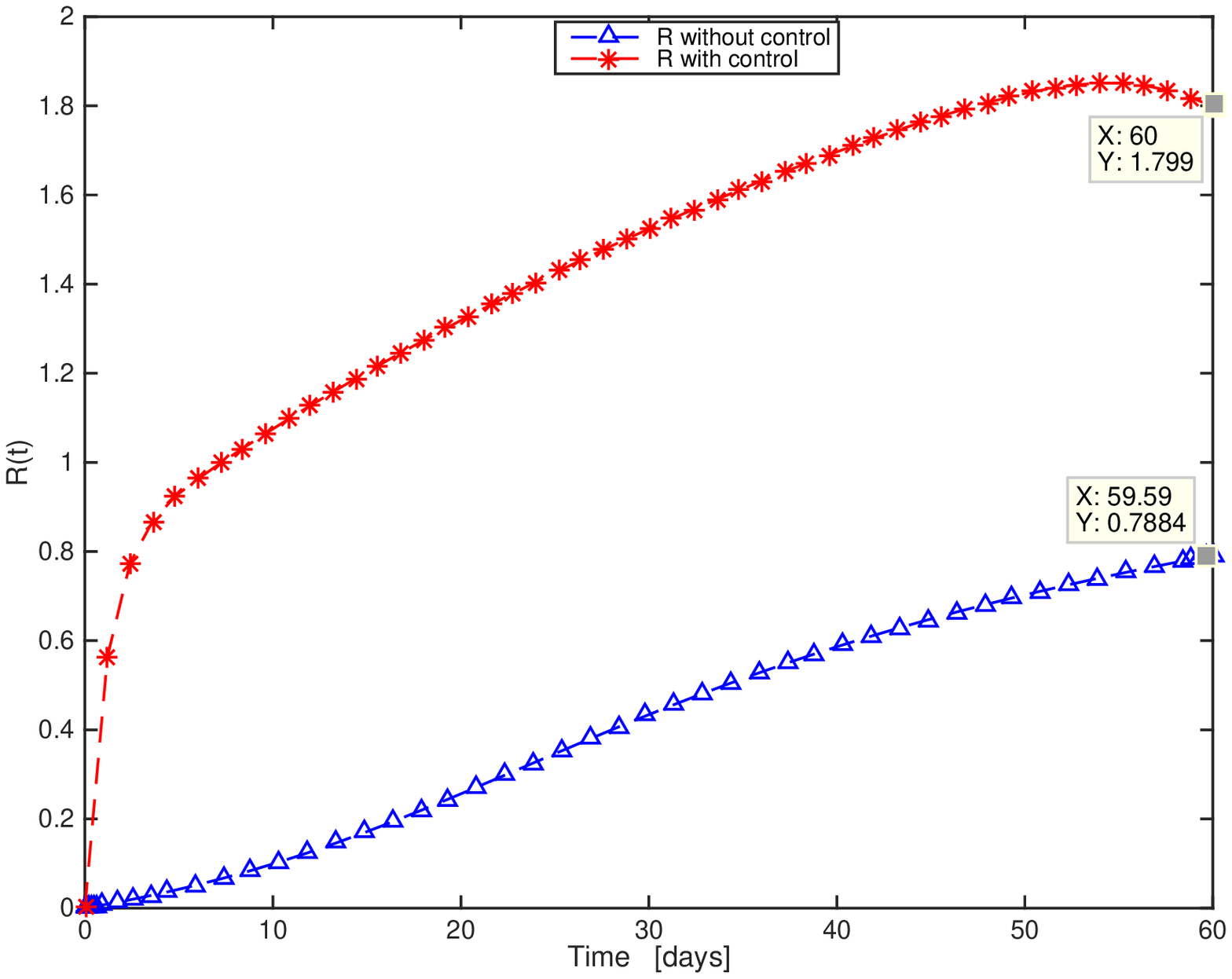}}\\
\subfloat[Infected $I(t)$]{
\label{cntrlBD_I}\includegraphics[width=0.50\textwidth]{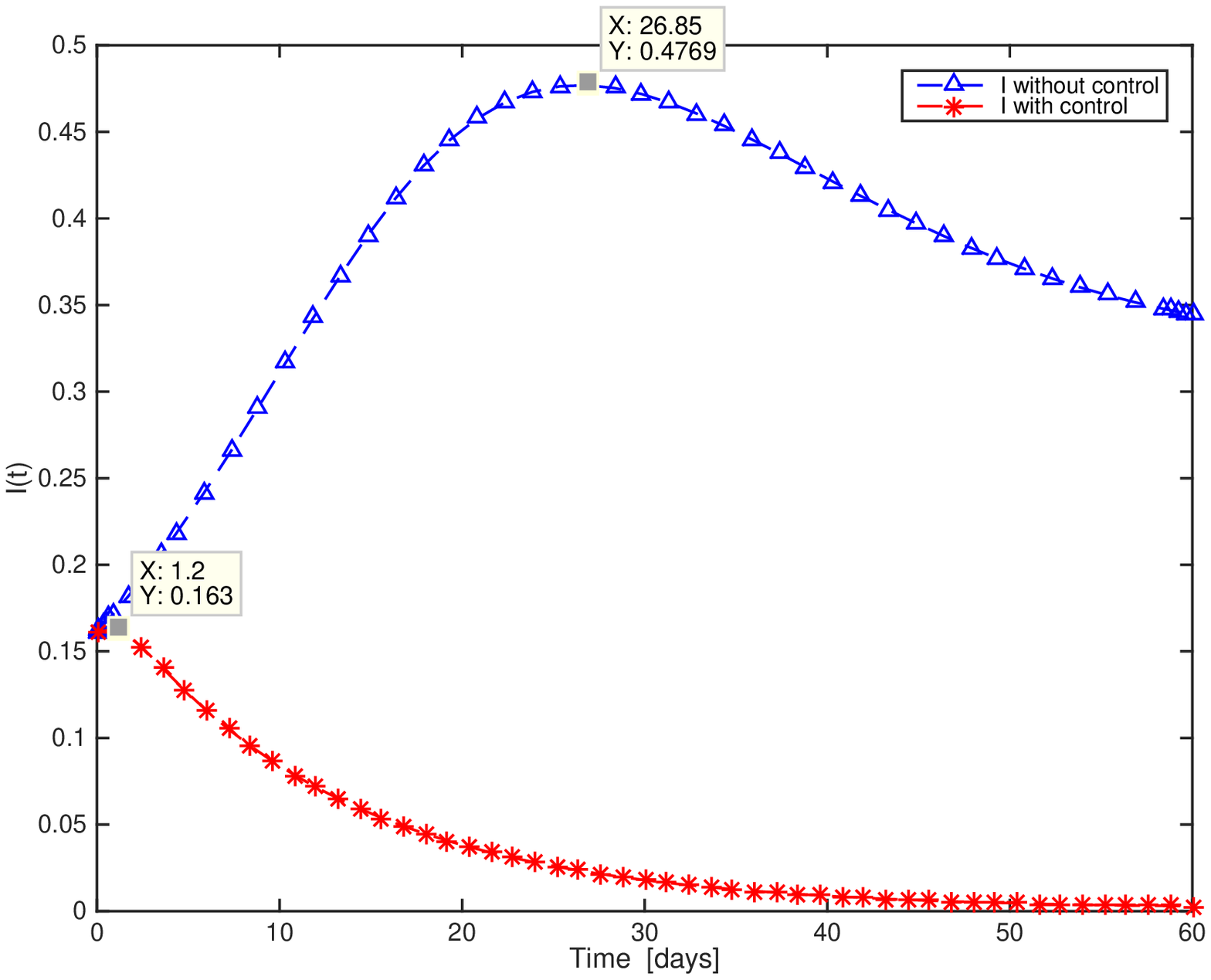}}
\caption{Comparison between the curves of individuals in case of optimal
control \eqref{SIR_BD_control}--\eqref{cost_func} with $\tau = 0.02$
\emph{versus} without control.}
\label{fig6}
\end{figure}
Figure~\ref{fig6} shows the significant difference in the number of susceptible,
recovered, and infected individuals with and without control.
In Figure~\ref{cntrlBD_S}, we see that the number of susceptible $S$,
in case of optimal control, decreases faster during the vaccination campaign.
It reaches 34.78\% at the end of the campaign against 44.12\% in the absence
of optimal control. Figure~\ref{cntrlBD_R} shows that the number of recovered
individuals increases rapidly. The number $R(t_{end})$ increases more rapidly 
in case of control than without control. In Figure~\ref{cntrlBD_I}, the 
time-dependent curve of infected individuals shows that the peak of the curve 
of infected individuals is less important in case of control. In fact, 
the maximum value on the infected curve $I$ under optimal control is 16.3\%, 
against 47.69\% without any control (see Figure~\ref{cntrlBD_I}). The other 
important effect of control, which we can see in the same curve, is the period 
of infection, which is less important in case of control of the virus. The value 
of the period of infection is $46$ days in case of optimal control, 
versus more than $60$ days without vaccination. This shows the efficiency 
of vaccination in controlling Ebola. 
\begin{figure}[ht!]
\centering
\includegraphics[scale=0.45]{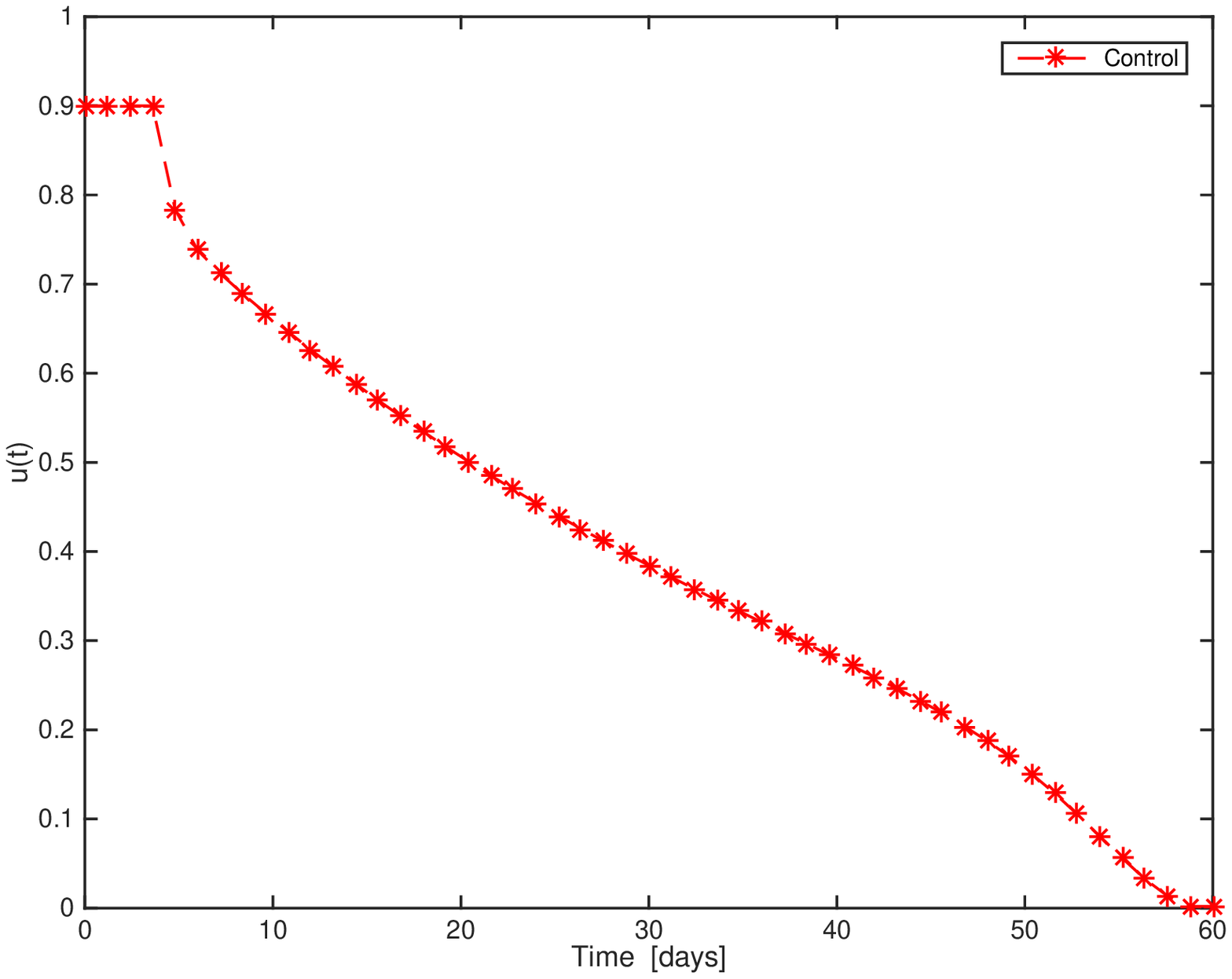}
\caption{Optimal control function $u(t)$
for problem \eqref{SIR_BD_control}--\eqref{cost_func}
with initial conditions \eqref{eq:ic_induced}, $t \in [0, t_{end}]$,
$t_{end} = 60$ days, and $\tau = 0.02$.\label{cntrlBD_u}}
\end{figure}
\begin{figure}[ht!]
\centering
\includegraphics[scale=0.45]{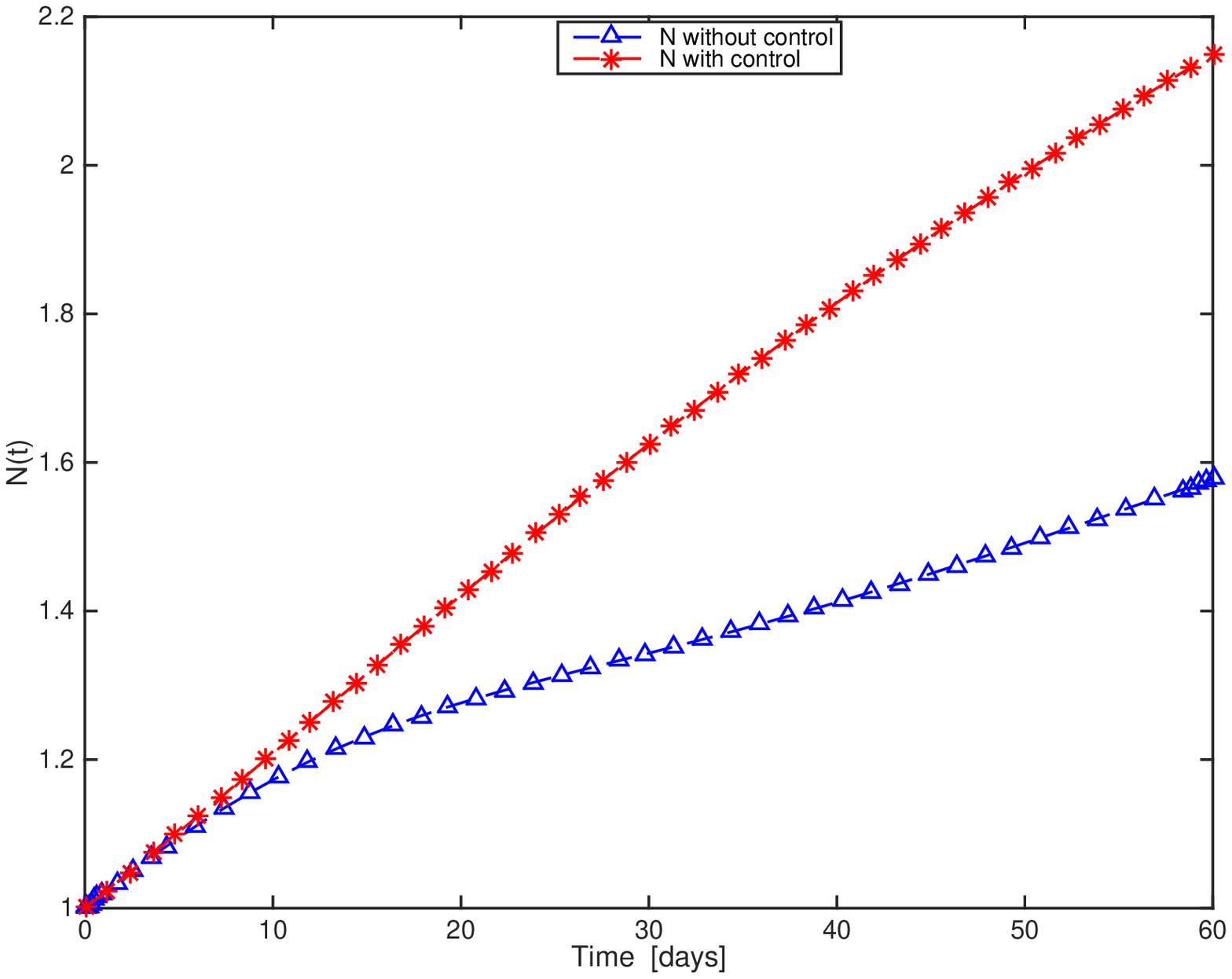}
\caption{Total population $N(t)=S(t)+I(t)+R(t)$, 
in case of optimal control \eqref{SIR_BD_control}--\eqref{cost_func}
\emph{versus} without control \eqref{SIR_BD2}.\label{cntrlBD_N_induced}}
\end{figure}
In conclusion, one can say that Figure~\ref{fig6} shows 
the effectiveness of optimal vaccination in controlling Ebola.
Figure~\ref{cntrlBD_u} gives a representation 
of the optimal control $u(t)$; while Figure~\ref{cntrlBD_N_induced} 
shows the evolution of the number of total population 
$N(t)=S(t)+I(t)+R(t)$ over time. We see that the total number 
of population is bigger in case of vaccination (less people dying). 


\section{Conclusion}
\label{sec:conc}

Mathematical modelling of the detection of a virulent virus such Ebola
is a powerful tool to understand the dynamics of the propagation of the virus
in a population. The main aim is to provide useful predictions about
the potential transmission of the virus. The important step after modelling
is to study the properties of the system of equations that describes
the propagation of the virus. In this work, we analysed 
a SIR model with vital dynamics for the early detection of Ebola virus 
in West Africa, by adding demographic effects and an induced death rate,
in order to discuss when the model makes sense mathematically and to study
the information provided by the model. We simulated the model in the case
of a basic reproduction number $R_0>1$, which describes the current situation
of Ebola virus in Guinea. We studied the equilibria. 
The system of equations of the model was solved numerically and 
the numerical simulations confirmed the theoretical analysis of the equilibria 
for the model. Finally, we controlled the propagation of the virus 
by minimizing the number of infected individuals and the cost of vaccination
and showing the importance of optimal control.  


\section*{Acknowledgements}

This research was partially supported by the Portuguese Foundation
for Science and Technology (FCT) through project UID/MAT/04106/2013
of the Center for Research and Development in Mathematics
and Applications (CIDMA) and within project TOCCATA,
ref. PTDC/EEI-AUT/2933/2014. 

The authors are very grateful 
to a referee for valuable remarks and comments, which
significantly contributed to the quality of the paper.



\bigskip


\end{document}